\documentstyle[11pt,newpasp,twoside,epsf]{article}
\def\edcomment#1{\iffalse\marginpar{\raggedright\sl#1\/}\else\relax\fi}
\reversemarginpar

\begin{document}
\title{Emission and absorption in Narrow-Line Seyfert 1 Galaxies}
\author{Anca Constantin \& Joseph C. Shields}
\affil{Department of Physics and Astronomy, Ohio University, Athens, OH 45701}

\begin{abstract}

The emission and absorption attributes of the UV-blue nuclear spectra
of the Narrow-Line Seyfert 1 (NLS1) galaxies are analyzed based on
high quality archival HST observations.  Measurements from composite
spectra as well as from individual sources reveal differences from the
more general AGN samples: NLS1s have steeper (redder) UV-blue
continua.  Objects with UV line absorption show redder spectra,
suggesting that internal dust is important in modifying the continuum
shapes.  A strong relationship is found between the slopes of the
power-laws that best fit the UV-blue continua and the luminosities:
the more luminous sources have bluer SEDs.  This trend is possibly
attributed to a luminosity-dependent inner geometry of the obscuring
(dusty) material.  


\end{abstract}

\section{Introduction}

NLS1s are interesting objects due to their extreme continuum and
emission-line properties.  Their fame as peculiar AGNs is mostly built
on analysis of either individual objects or, with the exception of
X-ray observations, small samples.  Detailed investigations of the
NLS1 UV/blue spectral properties are particularly limited (e.g.,
Rodriguez-Pascual, Mas-Hesse \& Santos-Lleo 1997, Kuraszkiewicz \&
Wilkes 2000); however, the mounting number of high-quality HST spectra
of these sources allows now for a better definition of their spectral
properties in general, and of their UV emission in particular.

This project attempts a comprehensive UV-optical spectral
characterization of the typical NLS1 galaxy.  The emission and
absorption characteristics of a sample of 22 NLS1-like objects are
analyzed by employing measurements in both individual and composite
spectra (see Constantin \& Shields 2003 for details on the sample
definition and data analysis).

\begin{figure}
\plotone{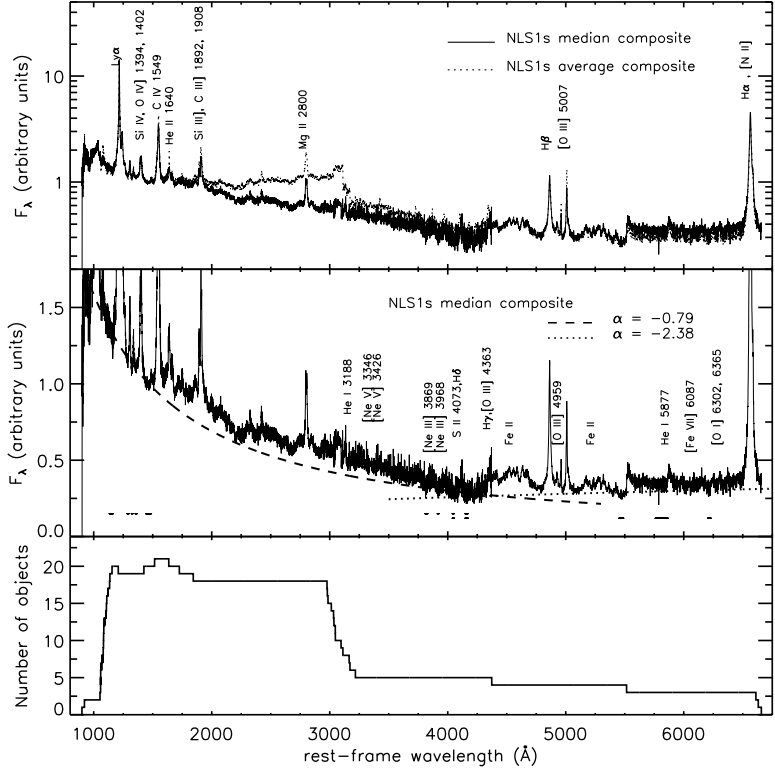}
\caption{{\it Top panel}: NLS1 composite spectrum plotted as
log(F$_{\lambda}$) vs. rest-frame wavelength, with the principal
emission features identified.  The flux has been normalized to unit
mean flux over the wavelength range 1430 \AA -- 1470 \AA.  {\it Middle
panel}: The broken power-law ($F_{\nu} \propto \nu^{\alpha}$) that
best fits the continuum is overplotted on the median composite, which
is plotted here on a linear scale and zoomed near the continuum level
for a better visualization of the weak features. {\it Bottom panel}:
Number of NLS1s contributing to the composite as a function of
rest-frame wavelength.  Only very few objects span the whole spectral
range; the UV-blue wavelengths are however well represented by the
sample.}
\end{figure}

Figure 1 shows the NLS1 composite spectra constructed with this (HST
archival) NLS1 data.  The power-law
that best fits the UV-blue continuum in the NLS1 median composite has
an index $\alpha = -0.79$ that falls among the steepest values found
in other AGN composite measurements (e.g., $\alpha = -0.44$ in the
SDSS sample, Vanden Berk et al. 2001, and $\alpha = -0.36$ for the
LBQS sample, Francis et al. 1991).  Understanding the origin of the
NLS1 continuum redness is important as this may be related to the
overall peculiar nature of these objects.


\section{What causes the redness of the NLS1s?}

A first clue to what may trigger the relatively steep spectral slope
in the NLS1 composite comes from the relation found between the
spectral slope $\alpha$ and the luminosity (Figure 2, left panel):
steeper slopes are measured in the lower luminosity sources.  Since
NLS1s have typically lower luminosities than those of the sources
employed in most other quasar composites, the $\alpha - L$ correlation
might explain the trend toward redder continua in these objects.

Independent evidence suggests further that the luminosity dependence
of the spectral index is mostly attributable to a luminosity-dependent
reddening.  In almost half the objects in this sample, absorption near
the systemic velocity of the host galaxy is present.  Moreover, the
spectral indices measured in the subsamples with and without UV
resonance line-absorption are in median --1.34 and --0.73,
respectively, suggesting that dust, which is expected to accompany the
absorbing gas, plays an important role in modifying the continuum
shape in these objects.  The median logarithmic luminosities for the
absorbed and unabsorbed subsamples are 29.01 and 29.60 respectively,
thus directly linking the presence of absorption with luminosity.


How can dust shape the QSO/NLS1 continuum?  A natural expectation is
that the circumnuclear interstellar medium assumes a disk-like
geometry with a luminosity-independent scale height, and a sublimation
radius $R_s$ that scales with the intrinsic luminosity $L$ of the
central source ($R_s \propto L^{1/2}_{\rm bolometric}$, a $receding\
torus$-type picture, Lawrence 1991).  In this simple scenario, a
sample of quasars with a random distribution of intrinsic bolometric
luminosities, inclination angles, and power-law indices of their
intrinsic SEDs, will display UV-blue spectral indices that correlate
strongly with the observed (reddened) luminosity.  Figure 2, right
panel, illustrates the results of simulating such dust extinction in
an ensemble of quasars; the torus properties (particle density $n_H$,
torus height $h$, extinction law $R_{\lambda}$) are considered the
same for all objects.  It is thus apparent that this framework offers
a plausible explanation for the observed $\alpha - L$ trend that is
also present in other, more heterogeneous AGN samples.  Interestingly,
the resulting numbers of absorbed sources predicted by this model
appear consistent with recent x-ray observations of the extragalactic
point source population.


Analysis of soft X-ray data could further test the nature of the dusty
absorber in these sources; a low ionization gas would produce strong
absorption, and therefore flatter observed soft X-ray spectra, while a
``warm'' (high ionization) absorber would imply that steeper soft
X-ray continua should be measured (e.g., Grupe et al. 1998).  A
comparison of the median values of the ROSAT spectral indices ($\Gamma
= 1 - \alpha$) for the absorbed and unabsorbed objects shows however
only a slight difference (3.4 and 3.1, respectively) suggesting that
the absorbing material may possess a range of properties.

\begin{figure}
\plotone{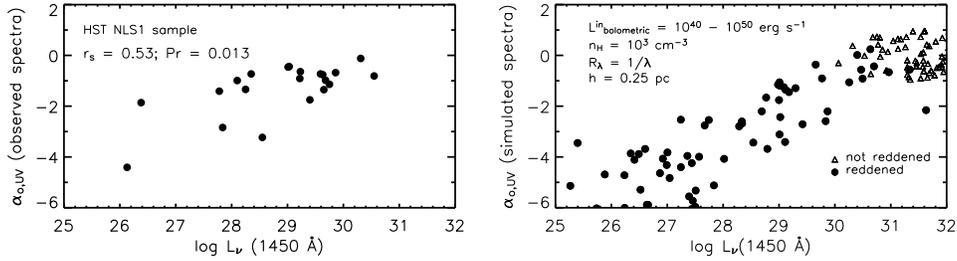}
\caption{{\it Left panel:} Spectral indices measured in the HST NLS1
sample plotted vs.  1450\AA\ luminosity, with $L_{\nu}$ expressed in
erg s$^{-1}$ Hz$^{-1}$.  The Spearman rank coefficient and the
probability of the correlation happening by chance are indicated.  The
error bars in both directions are smaller than the symbol size, and
therefore not indicated. {\it Right panel:} UV-optical spectral
indices plotted vs.  $L_{\nu}$ (1450\AA) luminosities calculated for a
simulated ensemble of quasars; the $\alpha -L$ trend is recovered by
simply employing a luminosity-dependent reddening by dust.}
\end{figure}

\section{Conclusions}

An analysis of all publicly available spectra of NLS1 galaxies in the
HST archive is presented here in an attempt to determine the UV-blue
spectral properties of these sources.  It is found that the NLS1s have
redder continua than that measured in other more general AGN samples.
In this sample, the spectral slope correlates strongly with the
luminosity indicating that the redness of the NLS1s is related to the
low luminosity of these objects.  Moreover, the apparent connection
between the UV resonance absorption lines and luminosity suggests that
the steep slopes measured in these source are due at least partially
to reddening.  Simple simulations show that a luminosity dependence of
the solid angle covered by the dust (as seen by the central source)
potentially explains the $\alpha - L$ trend.  The ionization state of
the absorbing material and its relationship to the accretion source
remain however ambiguous.


\begin{references}
\reference Constantin A., \& Shields, J. C., 2003, \pasp, 115, 592
\reference Grupe, D., Beuermann, K., Mannheim, K., \& Thomas, H.-C., 
1998, A\&A, 350, 805
\reference Francis, P. J., Hewett, P. C., Foltz, C. B., Chaffee, F. H., 
Weymann, R. J., \& Morris, S. L. 1991, \apj, 373, 465
\reference Kuraszkiewicz, J., \& Wilkes, B. J., 2000, \apj, 542, 692
\reference Lawrence, A., 1991, \mnras, 252, 586
\reference Rodriguez-Pascual, P. M., Mas-Hesse, J. M., \& Santos-Lleo, M., 
1997, A\&A, 327, 72
\reference Vanden Berk, D. E., et al., 2001, \aj, 122, 549
\end{references}
\end{document}